# CORRECT USE OF THE LIFSHITZ-SLYOSOV-WAGNER EXPRESSION FOR THE CALCULATION OF THE AVERAGE RADIUS OF AN OIL-IN-WATER (O/W) EMULSION SUBJECT TO FLOCCULATION AND COALESCENCE

Kareem Rahn-Chique y German Urbina-Villalba*

*SUMMARY*

*A novel theoretical expression for the estimation of the temporal variation of the average radius of a nanoemulsion subject to flocculation, coalescence and Ostwald ripening is proposed. It is based on the experimental evaluation of the mixed flocculation-coalescence rate. The predictions of the theory are contrasted with experimental results corresponding to a set of dodecane-in-water nanoemulsions stabilized with sodium dodecyl sulphate. A satisfactory agreement is found whenever the optical cross section of the aggregates is conveniently represented.*

*RESUMEN*

*Se propone una nueva expresión para la estimación de la variación temporal del radio promedio de una nanoemulsión sujeta a floculación, coalescencia y maduración de Ostwald. Ella está basada en la evaluación experimental de la tasa de floculación y coalescencia. Las predicciones de la teoría se contrastan con resultados experimentales correspondientes a un conjunto de nanoemulsiones de dodecano en agua estabilizadas con dodecil sulfato de sodio. Se encuentra un acuerdo satisfactorio siempre que la sección transversal óptica de los agregados sea convenientemente representada.*

**Kareem Rahn-Chique**. Licenciado en Química de la Universidad Central de Venezuela. Profesional Asociado a la Investigación (PAI-D2) del Instituto Venezolano de Investigaciones Científicas (IVIC). Centro de Estudios Interdisciplinarios de la Física. Caracas, Venezuela. Email: krahn@ivic.gob.ve.

**German Urbina-Villalba**. Doctor en Ciencias mención Química de la Universidad Central de Venezuela. Investigador Titular del IVIC. Jefe del Laboratorio de Fisicoquímica de Coloides. Centro de Estudios Interdisciplinarios de la Física. Caracas, Venezuela, Email: guv@ivic.gob.ve

1. INTRODUCTION

According to the Laplace equation (Evans and Wennerström, 1994), the internal pressure of a drop of oil suspended in water, is directly proportional to its interfacial tension ($\gamma$), and inversely proportional to the radius of the drop ($R_i$). The tension causes a difference between the chemical potential of the molecules of oil inside the drop and the ones belonging to an unbounded bulk oil phase. This difference ($\Delta\mu$) is equal to [Kabalnov, 1991]:

$$\Delta\mu = \frac{2\gamma V_M}{R_i} \qquad (1)$$

Where $V_M$ is the molar volume of the oil. According to Eq. (1) the excess chemical potential is positive, which means that a particle will be always dissolving when it is in contact with an aqueous bulk phase. Furthermore the derivative of the excess chemical potential is negative:

$$\frac{d\Delta\mu}{dR} = -\frac{2\gamma V_M}{R_i^2} \qquad (2)$$

This means that an ensemble of particles of different sizes cannot be in equilibrium with each other [Kabalnov, 2001]. As a result, the particles exchange molecules of oil through the solvent in a process referred to as Ostwald ripening. The theories of ripening start from the Kelvin equation:

$$C(R_i) = C_\infty \exp\left(\left(\frac{1}{R_i}\right)\frac{2\gamma V_m}{\tilde{R}T}\right) = C_\infty \exp\left(\frac{\alpha}{R_i}\right) \approx C_\infty\left(1 + \frac{\alpha}{R_i}\right) \qquad (3)$$

Here: $\tilde{R}$ is the universal gas constant, $T$ the absolute temperature, $R_i$ the radius of a drop, and $C(\infty)$ the solubility of the molecules in the presence of a planar oil/water (o/w) interface.

Thus, the difference between the aqueous solubility of a slab of oil in contact with water and the one of a drop of oil submerged in water ($C(R_i)$), depends on the quotient between the capillary length of the oil ($\alpha$) defined as:

$$\alpha = 2\gamma V_M / \tilde{R} T \qquad (4)$$

and the radius of the particle. According to the theory of Lifthitz, Slyosov, and Wagner (LSW theory) (Lifshitz and Slesov, 1959; Lifshitz and Slyosov, 1961; Wagner, 1961), the radius of a particle changes with time according to:

$$\frac{dR_i}{dt} = \frac{D_m}{R_i}\left(\Delta - \frac{\alpha}{R_i}\right) \qquad (5)$$

Where $D_m$ is the diffusion coefficient of the oil, and $\Delta$ is the supersaturation of the solution, $\Delta = C(R_i) - C(\infty)$. For each value of $\Delta$ there exists a critical radius ($R_c = \alpha/\Delta$) at which the particle is in equilibrium with the solution. Otherwise, the drop either grows ($R_i > R_c$) or dissolves ($R_i < R_c$). Eventually, a "stationary regime" is attained, characterized by a self-similar drop size distribution. At this time, the ripening rate ($V_{OR}$) can be quantified in terms of a linear increase of the cube of the critical radius ($R_c$) of the colloid as a function of time:

$$V_{OR} = dR_c^3/dt = 4\alpha D_m C(\infty)/9 \qquad (6)$$

Finsy (2004) demonstrated that the critical radius ($R_c$) of the dispersion is equal to its number average radius ($R_a$):

$$R_c = R_a = \frac{1}{N_T}\sum_k R_k \qquad (7)$$

According to our simulations (Urbina-Villalba et al., 2009b; Urbina-Villalba et al. 2012; Urbina-Villalba, 2014) the stationary regime results from two opposite processes: the exchange of oil molecules alone, which leads to a *decrease* of the average radius, and the elimination of particles (by dissolution or coalescence), which favors the *increase* of the average radius). The result is a saw-tooth variation of $R_a$ which was recently studied experimentally by Nazarzadeh et al. (Nazarzadeh, 2013). Hence, in the absence of other destabilization mechanisms, the stationary regime approximately begins as soon as the first particle dissolves; a phenomenon that occurs much earlier in nano-emulsions than in macro-emulsions.

Integration of Eq. 6 between $t_0$ and t leads to:

$$R_c^3(t) = V_{OR}[t - t_0] + R_c^3(t_0) \qquad (8)$$

Where: $t_0$ corresponds to the initial time of the measurements. Equation (8) is invariably used for the experimental evaluation of the Ostwald ripening rate (see for example references: [Kabalnov et al., 1990; Taylor, 1995; Weiss et al., 1999; Izquierdo et al., 2002; Tadros et al., 2004; Sole et al., 2006; Sole et al., 2012; Nazarzadeh et al., 2013]). However, the average radius weighted in volume or area is commonly used for this purpose. Moreover, the vast majority of the equipments available for static and dynamic light scattering measure the hydrodynamic radius of the "particles". Hence, they are unable to distinguish between the radius of a big drop and the hydrodynamic radius of an aggregate of drops. This creates an uncertainty regarding the effect of flocculation and coalescence on the increase of the average radius of the emulsion. If aggregation occurs, the slope $dR_c^3/dt$ will not be the product of ripening alone, and yet it will be usually contrasted with Eq. (6).

## 2. AVERAGE RADIUS OF AN EMULSION SUBJECT TO FLOCCULATION AND COALESCENCE

Recently, a novel theoretical expression for the turbidity of an emulsion ($\tau$) as a function of time was deduced (Rahn-Chique et al., 2012a; Rahn-Chique et al., 2012b, Rahn-Chique et al., 2012c). According to this equation, and in the absence of mixed aggregates (Rahn-Chique et al., 2012a), the scattering of light results from: the original drops of the emulsion (primary drops), aggregates of primary drops, and larger spherical (secondary) drops resulting from the coalescence of primary drops:

$$\tau = n_1 \sigma_1 + \sum_{k=2}^{k_{max}} n_k \left[ x_a \sigma_{k,a} + x_s \sigma_{k,s} \right] \qquad (9)$$

Here $\sigma_{k,a}$ and $\sigma_{k,s}$ represent the optical cross sections of an aggregate of size $k$ and the one of a spherical drop of the same volume. In the original formalism, $x_a$ stands for the fraction of collisions that only leads to the flocculation of the drops, $x_s$ represents the fraction of collisions that results from their coalescence, and $n_k$ is the number density of aggregates of size $k$ existing in the dispersion at time t (Smoluchowski, 1917):

$$n_k(t) = \frac{n_0 (k_{FC} n_0 t)^{k-1}}{(1 + k_{FC} n_0 t)^{k+1}} \qquad (10)$$

In Eq. (10) $n_0$ is the total number of aggregates at time t = 0 ($n_0 = \sum n_k(t=0)$) and $k_{FC}$ is an average aggregation-coalescence rate. When the number of mixed aggregates is small: $x_s = (1-x_a)$. The values of $k_{FC}$ and $x_a$ are obtained fitting Eq. (9) to the experimental variation of the turbidity as a function of time.

If the average radius of an emulsion is solely the result of flocculation it can be calculated as:

$$R_a = R_{FC} = \sum_{k=1}^{k_{max}} \left[\frac{n_k}{n}\right] R_k \tag{11}$$

Where $R_k$ is the average radius of an aggregate composed of k primary particles, and $n$ the total number of aggregates per unit volume. The term in parenthesis corresponds to the probability of occurrence of an aggregate of size k at a given time. The value of $n$ is equal to:

$$n(t) = \sum_{k=1}^{k_{max}} n_k(t) = \frac{n_0}{1 + k_{FC} n_0 t} \tag{12}$$

Where $k_{FC} = k_F$. If an emulsion is subject to both flocculation and coalescence (FC), the average size of the aggregates results from the contributions of the "true" aggregates of the particles, and the bigger drops resulting from coalescence. According to our simulations (Urbina-Villalba et al., 2005; Urbina-Villalba, 2009a) Eq. (12) also applies to the mixed process of flocculation and coalescence whenever coalescence is much faster than flocculation. Thus, every time Eq. (9) is valid, Eq. (11) can be recast in the following form:

$$R_{FC} = \left[\frac{n_1}{n}\right] R_1 + \sum_{k=2}^{k_{max}} \left[\frac{n_k}{n}\right] \left[x_a R_{k,a} + x_s R_{k,s}\right] \tag{13}$$

where: $R_{k,a}$ corresponds to the average radius of the aggregates with k primary particles, and $R_{k,s}$ to the radius of a drop resulting from the coalescence of k primary particles (which should be equal to $R_{k,s} = \sqrt[3]{k}\, R_0$). Only in the case in which $R_{k,a} \approx R_{k,s}$, Eq. (13) can be easily evaluated:

$$R_{FC} = \left[\frac{n_1}{n}\right]R_1 + \sum_{k=2}^{k_{max}}\left[\frac{n_k}{n}\right]R_{k,a}[x_a + x_s] = \sum_{k=1}^{k_{max}}\left[\frac{n_k}{n}\right]R_k \qquad (14)$$

since: $x_s = (1 - x_a)$. Otherwise the dependence of $R_{k,a}$ on k is unknown due to the variety of conformations available for each size. Alternative approximate expressions can be forwarded based on Eq. (9) and the connection between the optical cross section of a *spherical* particle (σ) and its radius ($\sigma = Q_s \pi R^2$):

$$R_{FC} = n_1 (\sigma_1/Q_s \pi)^{1/2} + \sum_{k=2}^{k_{max}} n_k \left[ x_a \left(\sigma_{k,a}/Q_a \pi\right)^{1/2} + x_s \left(\sigma_{k,s}/Q_s \pi\right)^{1/2} \right] \qquad (15)$$

where Q is the scattering coefficient (Gregory, 2009). The value of $Q_s$ can be estimated using the program Mieplot (www.philiplaven.com) for the initial particle size. Alternatively, a k-dependence of the average radius of an aggregate composed of k primary drops of size $R_0$ can be extrapolated based on the variation of the average radius of coalescing drops:

$$R_{FC} = n_1 R_0 + \sum_{k=2}^{k_{max}} n_k \left[ x_a\, k^m + x_s\, k^n \right] R_0 \qquad (16)$$

where *m* and *n* are rational numbers. Hence: it is possible to calculate the variation of the average radius of an emulsion due to flocculation and coalescence alone, if the value of $k_{FC}$ can be obtained by adjusting Eq. (9) to the experimental data.

# 3. AVERAGE RADIUS OF AN EMULSION SUBJECT TO FLOCCULATION, COALESCENCE, *AND OSTWALD RIPENING*

From the previous sections it is clear that once the value of $k_{FC}$ has been evaluated by means of Eq. (9), equations (14), (15) or (16) can be employed to predict the effect of flocculation and coalescence on the temporal variation of the average radius. Instead, equation (8) can be used to predict the sole influence of Ostwald ripening as predicted by LSW. In fact, the change of the average number of *drops* ($n_d$) due to Ostwald ripening, has the same structure of Smoluchowski equation (Weers 1999; Urbina-Villalba et al., 2014):

$$n_d = \frac{n_0}{1 + k_{OR}\, n_0\, t} \tag{17}$$

Where:

$$k_{OR} = \frac{16\,\pi\,\alpha\,D_m\,C_\infty}{27\,\phi} \tag{18}$$

Here $\phi$ is the volume fraction of oil. Eq. (17) is similar to Eq. (12) and therefore, the evaluation of the rate constant by means of the change in the number of aggregates as a function of time does not identify the process of destabilization.

It is evident that flocculation, coalescence and ripening are not independent processes. This situation is clearly illustrated in the algorithm proposed by De Smet et al. (De Smet et al., 1997) to simulate the process of Ostwald ripening. Starting from Fick´s law, using the Kelvin´s equation, and assuming that the capillary length of the oil is substantially lower than the radii of the drops ($\alpha \ll R_i$), these authors demonstrated that the number of molecules of a drop of oil (i) suspended in water ($m_i$), changes in time according to:

$$\frac{dm_i}{dt} = 4\pi D_m C_\infty \alpha \left( \frac{R_i(t)}{R_c(t)} - 1 \right) \tag{19}$$

According to Eq. (19) the drops can increase or decrease their volume depending on the quotient between their particular radius and the critical radius of the emulsion. The amount of molecules transferred to/from a particle depends on the referred quotient. Hence, if the average (critical) radius of the emulsion increases due to a mechanism different from Ostwald ripening, the process of ripening should also be affected. As previously noted, the simulations suggest that the average radius of the emulsion only increases due to ripening when the total number of drops decreases by dissolution. Hence, if the average radius of the emulsion increases faster than predicted by LSW, the number of drops that fall below the critical radius increases. Consequently, the number of drops subject to dissolution should be substantially larger than the one predicted by LSW. This should promote a further increase in the average radius, and a larger Ostwald ripening rate.

Let us study the simplest scenario in which Oswald ripening occurs independently of flocculation and coalescence. In this case the average radius of the emulsion at time t would be equal to:

$$R(t) = R(t_0) + (\Delta R)_{OR} + (\Delta R)_{FC} \tag{20}$$

Here the symbol $(\Delta R)_p$ stands for the change of the average radius due to process "p":

$$(\Delta R)_p = R_p(t) - R_p(t_0) \tag{21}$$

We had previously deduced an explicit expression for flocculation and coalescence (FC) which allows the evaluation of $(\Delta R)_{FC}$ (Eqs. (11), and (14) – (16)). In the case of Ostwald ripening (OR), equation (8) leads to:

$$R_{OR}(t) = \sqrt[3]{V_{OR}[t-t_0] + R_{OR}^3(t_0)} \tag{22}$$

Notice that:

$$R(t_0) = R_{FC}(t_0) = R_{OR}(t_0) = R_0 \tag{23}$$

Hence:

$$R(t) = R_0 + \left(\sqrt[3]{V_{OR}[t-t_0] + R_0^3} - R_0\right) + \left(R_{FC}(t) - R_0\right) \tag{24}$$

$$R(t) = \sqrt[3]{V_{OR}[t-t_0] + R_0^3} + R_{FC}(t) - R_0 \tag{25}$$

Equation (25) has the correct limit ($R_0$) for $t = t_0$, and it's the most accurate expression that can be derived, assuming that FC and OR occur simultaneously but independently.

An interesting expression results if Eq. (6) is integrated between $t - \Delta t$ and $t$, and it is assumed that the process of flocculation and coalescence occurs much faster than the one of Ostwald ripening ($R_{OR}^3(t-\Delta t) = R_{FC}^3(t-\Delta t)$):

$$R_c(t) = \sqrt[3]{V_{OR}\Delta t + R_{FC}^3(t-\Delta t)} \tag{26}$$

Using Eq. (26) recursively in order to reproduce the total time elapsed (see Eq. (A.6) in Appendix A), and assuming that the FC contribution is independent from the one of OR:

$$R_a(t) = R_{FC}(t) + R_c(t) \qquad (27)$$

one obtains a very good approximation to the experimental data for the systems with the stronger variation of the average radius. However, while Eq. (26) is a reasonable approximation to the critical radius, Eq. (27) is <u>incorrect</u>, since it obliterates the fact that there is only one average radius for the system ($R_a(t) = R_c(t)$), and therefore, Eq. (26) already contains the effects of OR *and* FC (see Appendix A).

4. EXPERIMENTAL DETAILS

*4. 1. Materials*

N-dodecane ($C_{12}$, Merck 98%) was eluted twice through an alumina column prior to use. Sodium dodecyl sulphate SDS (Merck) was recrystallized from ethanol two times. Sodium chloride NaCl (Merck 99.5%), and isopentanol (IP, Scharlau Chemie 99%) were used as received. Millipore´s Simplicity water was employed (conductivity $< 1\,\mu$ S cm$^{-1}$ at 25ºC).

*4. 2. Dispersion Preparation and Characterization*

An equilibrated system of water + liquid crystal + oil with 10% wt SDS, 8 wt% NaCl, 6.5 wt% isopentanol (iP) and a weight fraction of oil $f_o = 0.80$ ($f_o + f_w = 1$, $\phi = 0.84$) was suddenly diluted with water at constant stirring until the final conditions were attained: 5 wt% SDS, 3.3 wt% isopentanol, and $f_o = 0.38$ ($\phi = 0.44$). This procedure allowed the synthesis of mother nano-emulsions (MN) with an average radius of 72.5 nm. An appropriate aliquot of MN was then diluted with a suitable aqueous solution containing W, SDS, NaCl and iP (W/SDS/NaCl/iP) in order to obtain systems with $f_o$ = 0.35, 0.30, 0.25, 0.20, 0.15, 0.10, and salt concentrations of 2 and 4 wt% NaCl. To avoid the risk of perturbation and/or contamination of a unique sample, each of these emulsions was divided into 15 vials and stored at 25 °C. These vials were used later to study the evolution of the emulsion during 6 hours. The whole procedure was repeated thrice with independently-prepared mother nano-emulsions. In all cases, the concentration of SDS and iP was kept fixed at 5 wt% and 3.3 wt%, respectively. An additional set of emulsions was prepared leaving a set of MN to evolve in time until their average radius reached R ~ 500 nm (mother macro-emulsion: MM). As before, aliquots of MM were diluted with W/SDS/NaCl/iP solution in order to prepare systems with the same physicochemical conditions of the nano-emulsions, but with a higher particle radius. The average size of the dispersions, and their drop size distribution (DSD) were measured using a LS 230 from Beckman-Coulter. Out of the total of 24 systems, 6 representative emulsions were selected to illustrate the general behavior found (Table 1).

## 4. 3. Evaluation of $k_{FC}$

At specific times an aliquot was taken from one the set of vials corresponding to each system. The aliquot was diluted with an aqueous solution of SDS in order to reach a volume fraction of oil equal to $\phi = 10^{-4}$ ([SDS] = 8 mM). Then, the value of the turbidity was measured using a Turner spectrophotometer (Fisher Scientific) at $\lambda$ = 800 nm (Rahn-Chique et al., 2012a, Rahn-Chique et al., 2012c). This procedure was repeated for 6 hours. When the whole set of measurements was complete, Eq. (9) was fitted to the experimental data of the turbidity ($\tau = 230 Abs$) using Mathematica 8.0.1.0.

The optical cross sections used in Eq. (9) are valid whenever:

$$C_{RGD} = (4\pi R/\lambda)(m-1) << 1 \tag{28}$$

Here, $\lambda$ is the wavelength of light in the liquid medium, and $m$ the relative refractive index between the particle and the solvent. In the case of a dodecane/water emulsion (m = 1.07), the values of $C_{RGD}$ corresponding to radii R = 50, 60, 70, 80, 100, 500 nm are: 0.07, 0.09, 0.10, 0.12, 0.15, 0.75, respectively. These values are reasonably low, guaranteeing errors in the optical cross sections of the drops lower than 10% within the Raleigh-Gans-Debye approximation (Kerker, 1969).

The values of $k_{FC}$, $t_{0,teo}$ (theoretical starting time of the aggregation process), and $x_a$ ($x_s = 1 - x_a$) were directly obtained from the fitting of Eq. (9) to the experimental data. The theoretical value of the radius is a parameter of the calculation which can be systematically varied ($R_{teo} = R_{exp} \pm \delta$) to optimize the fitting, and guarantee a value of $t_{0,teo}$ close to the experiment. The errors bars were calculated using the procedure described in (Rahn-Chique et al., 2012a; 2012c). The effect of buoyancy during these measurements is known to be negligible (Cruz-Barrios et al., 2014).

For the evaluation of $k_{FC}$, aliquots of the concentrated emulsion were taken periodically, and diluted. Hence, the number of particles per unit volume used in the determination of $k_{FC}$ corresponds to a dilute (d) system ($k_{FC,d}$) and not to the actual, concentrated (c) emulsion ($k_{FC,c}$) under study. Simple arithmetic shows that $k_{FC,d}$ has to be multiplied by the dilution factor in order to obtain $k_{FC,c}$. If an aliquot of volume $V_c$ is removed from a concentrated emulsion with an aggregate density $n_c$, and diluted with aqueous solution until reaching a final volume $V_d$, the new aggregate density $n_d$ fulfills the following relationship:

$$n_d(t) V_d = n_c(t) V_c \tag{29}$$

This allows the definition of a dilution factor $f_d$:

$$n_d(t) = n_c(t)\left(\frac{V_c}{V_d}\right) = n_c(t) f_d \tag{30}$$

$$f_d = n_c(t)/n_d(t) \tag{31}$$

Using the expression of Smoluchowski for a total number of aggregates at time t:

$$\frac{n_{0,d}}{1 + k_{FC,d}\, n_{0,d}\, t} = \left(\frac{1}{f_d}\right) \frac{n_{0,c}}{1 + k_{FC,c}\, n_{0,c}\, t} \tag{32}$$

the following equality is obtained:

$$n_{0,d}\left(1 + k_{FC,c}\, n_{0,c}\, t\right) = \left(\frac{1}{f_d}\right) n_{0,c}\left(1 + k_{FC,d}\, n_{0,d}\, t\right) \tag{33}$$

But for any time t, and in particular for t = 0:

$$f_d = n_c(0)/n_d(0) = n_{0,c}/n_{0,d} \tag{34}$$

Therefore:

$$k_{FC,c}\, n_{0,c} = k_{FC,d}\, n_{0,d} \tag{35}$$

which means that the value of $k_{FC}$ obtained from the turbidity measurements of the dilute systems has to be divided by the dilution factor in order to get the rate of flocculation and coalescence of the actual concentrated emulsions:

$$k_{FC,c} = k_{FC,d} \frac{n_{0,d}}{n_{0,c}} = k_{FC,d} \frac{1}{f_d} \qquad (36)$$

*4.4. Prediction of the average radius*

The variation of the average radius of the emulsions as a function of time was estimated under different premises:

1) An order-of-magnitude estimation of the mixed flocculation/coalescence rate was obtained using the experimental radius of the emulsions to compute an approximate number of drops: n = $\phi/V_1$ (where $4/3\pi R^3$). A rough Smoluchoswki's rate ($k_S$) was calculated from the fitting of this data to Eq. (12) supposing that the number of aggregates was equal to the number of drops. Following, $R_a$ was computed using Eq. (11), and assuming $R_{k,a} \approx R_{k,s} = \sqrt[3]{k} R_0$.

2) Only Ostwald ripening occurs. The LSW radius was evaluated using Eq. (8). The value of $V_{OR}$ (2.6 x $10^{-28}$ m$^3$ s$^{-1}$) was estimated from Eqs. (4) and (6) using $\gamma$ = 1.1 mN m$^{-1}$, $C(\infty)$ = 5.4 x $10^{-9}$ cm$^3$ cm$^{-3}$, $V_M$ = 2.3 x $10^{-4}$ m$^3$ mol$^{-1}$, $D_m$ = 5.4 x $10^{-10}$ m$^2$ s$^{-1}$ (Sakai et al., 2002).

3) Only flocculation and coalescence occur. The procedure described in section 4.3 was used to calculate $k_{FC}$. The value of $R_a$ was estimated using Eq. (14).

4) Flocculation, coalescence and Ostwald ripening occur. The procedure described in section 4.3 was used to calculate $k_{FC}$. Following, Eqs. (25) was employed to estimate $R_a$.

5. RESULTS AND DISCUSSION

Depending on the relative importance of flocculation and coalescence with respect to Oswald ripening, there are at least three possible situations:

1) The average radius of the emulsion is substantially higher than the predictions of Eqs. (25) using Eq. (14) to approximate $R_{FC}$. This is illustrated by systems A and B of Tables 1 and 2 (curve FCOR in Figs. 1(a) and 1(b)).

2) Only flocculation and coalescence (FC) occur without a sensible contribution of ripening (systems C and D, Figs. 1(c) and 1(d)). In this case, the forecast of Ostwald ripening (OR) according to LSW is too low, and the prediction of Eq. (25), FCOR coincides with the one of FC. Both FCOR and FC lie very close to the experimental data.

3) Only ripening takes place (systems E and F, Figs. 1(e) and 1(f)). The curves corresponding to FC and FCOR largely surpass the experimental data. Instead, the LSW prediction reproduces very well the experimental measurements. In fact, only in the systems in which the initial average radius is large (showing the slowest variation of the average radius with time) the prevalence of Ostwald ripening is observed.

Fig 2 illustrates the effect of the initial time of measurement, the k-dependence of the cross section on the average radius of the aggregates (Eq. (16)), and the magnitude of Q ($Q = Q_s = Q_a$) in Eq. (15), on the predictions of Eq. (25) for systems A and B. It is clear that in these cases, the aggregates formed appear to be more linear than globular ($R_{k,a} = k^m R_0$, $m \approx 0.75$). As a consequence, the optical scattering coefficient that fits the experimental data lies approximately between one tenth (Q = 0.0025) and one third (Q = 0.11) of the values predicted by Mie theory (0.022 and 0.30, respectively) for *spherical drops* with the same initial average radius ($R_0$) of the emulsions corresponding to systems A and B. Thus, the discrepancy between the predictions of Eq. (25) and the experiment in the most unstable (concentrated) systems could possibly be adjudicated to our inability to represent the average radius of the floccules accurately.

## 6. CONCLUSION

A novel methodology able to predict the variation of the average radius of an emulsion as a function of time during a period of –at least- 6 hours is proposed. The procedure allows discriminating the relative importance of the processes of flocculation/coalescence with respect to Ostwald ripening on the temporal change of the average radius of the emulsions. A satisfactory agreement between theory and experiment depends on the soundness of the theoretical expression used to represent the dependence of the average radius of the clusters with the number of individual drops.

## 7. APPENDIX A

If it is assumed that:

$$\frac{dR}{dt} = \left(\frac{dR_{OR}}{dt}\right) + \left(\frac{dR_{FC}}{dt}\right) \tag{A.1}$$

and:

$$V_{OR} = dR^3/dt = 3R^2 \frac{dR}{dt} = 4\alpha D_m C(\infty)/9 \qquad (A.2)$$

The second term on the right hand side of Eq. (A.1) can be integrated, but the first term leads to an unsolvable integral due to our lack of knowledge on the variation of the radius with t.

$$\int \frac{dR}{dt} = V_{OR} \int \frac{dt}{R^2(t)} \qquad (A.3)$$

If on the other hand, it is supposed that:

$$\frac{dR^3}{dt} = \left(\frac{dR_{OR}^3}{dt}\right) + \left(\frac{dR_{FC}^3}{dt}\right) \qquad (A.4)$$

Integration yields:

$$R^3(t) - R^3(t_0) = V_{OR}[t - t_0] + R_{FC}^3(t) - R_{FC}^3(t_0) \qquad (A.5)$$

Simplifying:

$$R^3(t) = V_{OR}[t - t_0] + R_{FC}^3(t) \qquad (A.6)$$

This approximation leads to the correct limit for t = $t_0$. Moreover, the same *ansatz* leads to:

$$3R^2 \frac{dR}{dt} = 3R_{OR}^2 \left(\frac{dR_{OR}}{dt}\right) + 3R_{FC}^2 \left(\frac{dR_{FC}}{dt}\right) \qquad (A.7)$$

Or what is equivalent:

$$\frac{dR}{dt} = \frac{R_{OR}^2}{R^2}\left(\frac{dR_{OR}}{dt}\right) + \frac{R_{FC}^2}{R^2}\left(\frac{dR_{FC}}{dt}\right) \qquad (A.8)$$

Thus, the total first derivative of the radius is a linear combination of the contributions of ripening and flocculation-coalescence weighted on their relative importance.

FIGURE CAPTIONS

Fig 1. Comparison between the experimental data and the predictions of average radius as a function of time for systems A to E (Table 1). The values of the mixed flocculation/coalescence rates are given in Table 2. S, OR, FC, and FCOR, stand for the predictions of the theory for Smoluchowski (Eq. (12)), Ostwald ripening according to LSW (Eq. (8)), Flocculation and Coalescence (Eq. (14) with $R_{k,a} = R_{k,s} = \sqrt[3]{k} R_0$), and all processes combined (Eq. (25)).

Fig. 2. Predictions of Eq. (25) for systems A and B of Fig. 1 using different approximations for $R_{FC}$: (a) changing the initial time of measurement, (b) using different values of m for Eq. (16), and (c) adjusting the value of Q (= $Q_s$ = $Q_a$) in Eq. (15).

TABLE CAPTIONS

Table 1: Composition of the emulsions studied. Weight fraction of oil ($f_o$), salt concentration, initial radius, and particle density of the concentrated ($n_{0,c}$) and dilute ($n_{0,d}$) systems.

Table 2: Parameters obtained from the fitting of Ec. 9 to the experimental data. The errors listed in the Table result from the average of the errors of three independent measurements (Rahn-Chique, 2012a).

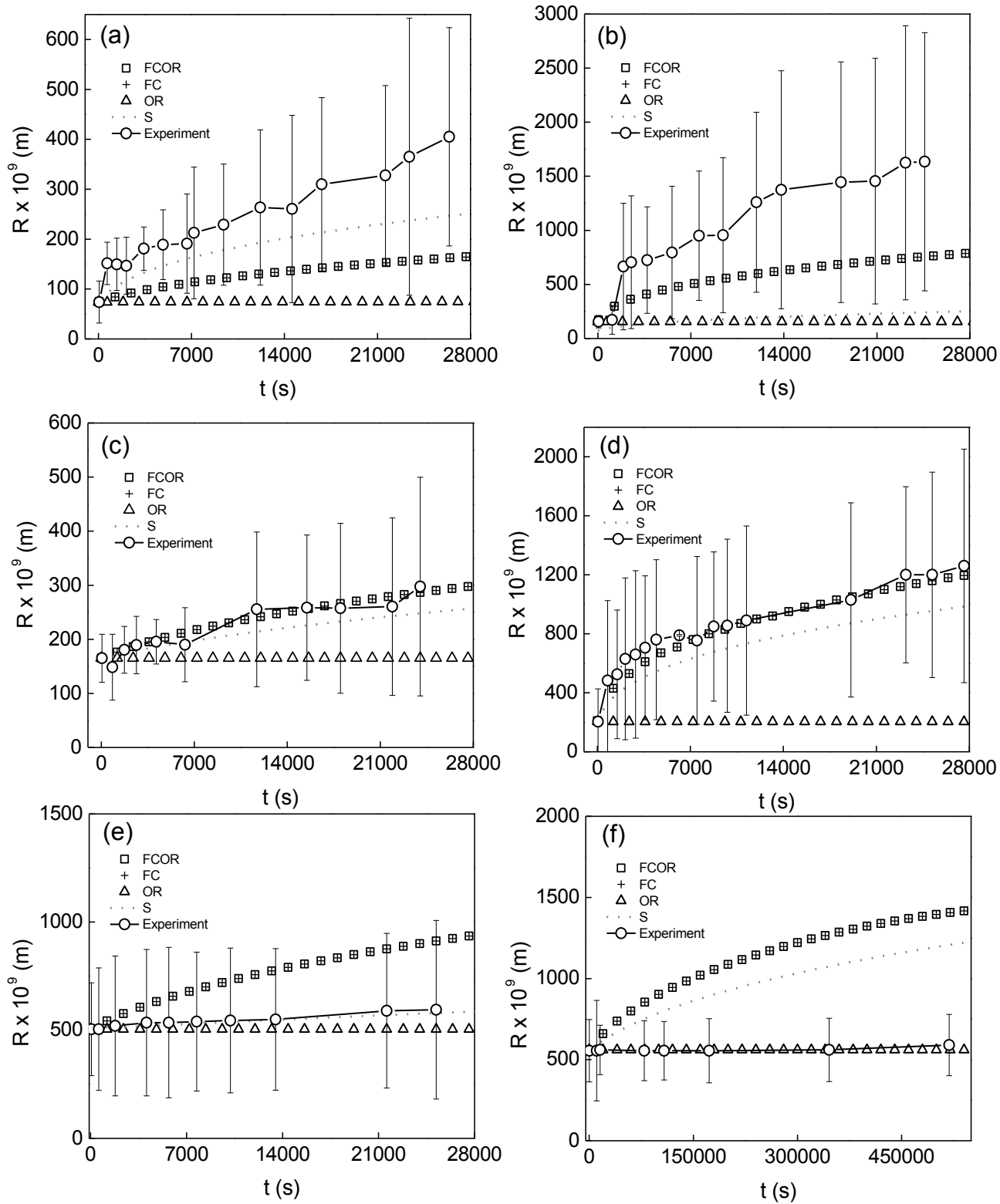

Fig. 1

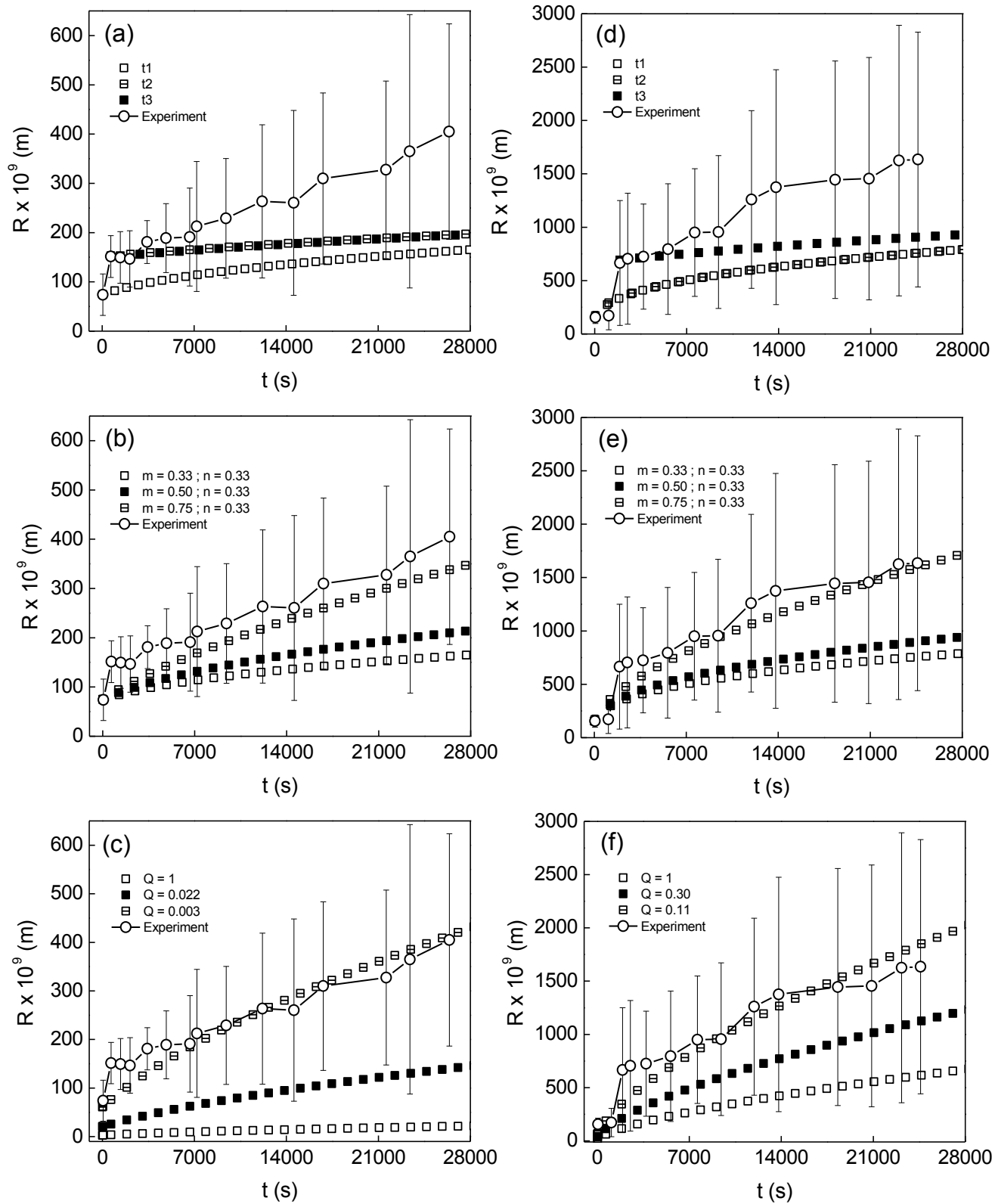

Fig. 2

Table 1

| ID | $f_o$ | % NaCl | $R_{0,exp}$ (nm) | $n_{0,c}$ (m$^{-3}$) | $n_{0,c}$ (m$^{-3}$) |
|---|---|---|---|---|---|
| A | 0.35 | 2 | 74 | $2.5 \times 10^{20}$ | $5.9 \times 10^{16}$ |
| B | 0.30 | 4 | 156 | $2.3 \times 10^{19}$ | $6.3 \times 10^{15}$ |
| C | 0.25 | 2 | 165 | $1.6 \times 10^{19}$ | $5.3 \times 10^{15}$ |
| D | 0.25 | 4 | 204 | $8.6 \times 10^{18}$ | $2.8 \times 10^{15}$ |
| E | 0.30 | 2 | 505 | $6.8 \times 10^{17}$ | $1.9 \times 10^{14}$ |
| F | 0.25 | 2 | 560 | $4.2 \times 10^{17}$ | $1.4 \times 10^{14}$ |

Table 2

| ID | $R_{0,teo}$ (nm) | $x_a$ | $k_{FC,d}$ (m$^3$ s$^{-1}$) | $k_{FC,c}$ (m$^3$ s$^{-1}$) | $t_0$ (s) |
|---|---|---|---|---|---|
| A | 90 | 0.51 ± 0.18 | (9.1 ± 1.6) x 10$^{-21}$ | (2.3 ± 0.4) x 10$^{-24}$ | 54.9 |
| B | 141 | 0.15 ± 0.16 | (1.3 ± 0.2) x 10$^{-18}$ | (2.8 ± 0.4) x 10$^{-22}$ | 18.6 |
| C | 173 | 0.97 ± 0.04 | (4 ± 6) x 10$^{-20}$ | (1.5 ± 2.1) x 10$^{-23}$ | 39.9 |
| D | 181 | 0.33 ± 0.04 | (2.0 ± 0.2) x 10$^{-18}$ | (7.0 ± 0.7) x 10$^{-22}$ | 14.5 |
| E | 520 | 1.5 ± 0.2 | (2 ± 12) x 10$^{-20}$ | (4 ± 30) x 10$^{-24}$ | 26.6 |
| F | 424 | 1.6 ± 0.2 | (1.6 ± 1.1) x 10$^{-20}$ | (1.1 ± 0.7) x 10$^{-23}$ | 8.1 |